\documentclass{aastex631}

\newcommand\Hunt{{\cite{2023A&A...673A.114H}}} 
\newcommand\Tarricq{{\cite{2022A&A...659A..59T}}}
\newcommand\Groeningen{{\cite{2023A&A...675A..68V}}}
\newcommand\Zerjal{{\cite{2023A&A...678A..75Z}}}
\newcommand\Bouvier{{\cite{1997A&A...323..139B}}}

\begin{document}

\title{Unveiling subarcsecond multiplicity in the Pleiades with Gaia multicolor photometry}

\author[0000-0001-5038-0089]{Dmitry Chulkov}
\affiliation{Institute of Astronomy of the Russian Academy of Sciences (INASAN) \\
119017, Pyatnitskaya st., 48, Moscow ; \href{mailto:chulkovd@gmail.com}{chulkovd@gmail.com}}



\begin{abstract}

The list of 409 probable cluster members down to $G=15^{\rm mag}$ ($m \gtrsim 0.5M_\odot$) is compiled for the two degree radius of the Pleiades, based on astrometric data from Gaia DR3 and the PPMXL catalog, along with several radial velocity surveys, including APOGEE and LAMOST. This approach allows for the inclusion of binary stars with unreliable Gaia solutions, thereby eliminating associated bias. Thus, the often-neglected 14 sources with Gaia two-parameter solutions are included. The subsequent analysis of color-magnitude and color-color diagrams exploits artifacts in Gaia photometric data, caused by the different field sizes used to measure fluxes in the $G$, $B_p$, and $R_p$ passbands, to reveal binary stars with subarcsecond angular separation. The findings are validated with prior high-resolution observations. Overall, $24 \pm 3$ cluster members with angular separation between 0.1 and 1 arcsec (13.5 to 135 AU projected distance) and mass ratio $q>0.5$ are deemed binary, indicating a binarity fraction of $6 \pm 1$\%.
\end{abstract}




\section{Introduction} \label{sec:intro}

\textcolor{black}{ Young clusters are important tracers of the structure and properties of their host galaxies \citep{2020SSRv..216...69A}, including our own Milky Way \citep{2023AJ....166..170J, 2023AstL...49..320B}. As coeval and chemically homogeneous groups, clusters are valuable for developing stellar evolution models \citep{2024NewAR..9901696C}. The topic of stellar multiplicity is intimately related to the study of open clusters, as most stars are born in dense stellar environments, and their dissolution gives rise to one of the binary formation mechanisms \citep{2010MNRAS.404.1835K}. Binarity complicates the determination of the initial and present-day mass functions, which are key characteristics of stellar populations \citep{2020SSRv..216...70L}. Despite major efforts in recent years, the statistical properties of binary stars and their variations in different environments remain controversial due to numerous selection effects associated with observational data \citep{2024PrPNP.13404083C}. The Gaia mission \citep{2016A&A...595A...1G} recently opened a new chapter both for open cluster and multiplicity studies \citep{2024NewAR..9801694E}, with a number of works dedicated to the characterization of multiplicity in the Galactic clusters (see references in Section \ref{multi}). The Pleiades has the highest number of members within 400 pc of the Sun \citep{2023MNRAS.526.4107P}, making it an important benchmark for astronomers. Its relative proximity allows for in-depth analysis that is currently unattainable for more distant objects. This paper begins with an inventory of the cluster members in Section \ref{Targets}, followed by the adoption of a theoretical isochrone used for mass and binary mass ratio calculations in Section \ref{isochrones}. The analysis of multiplicity, based on the interpretation of Gaia multicolor photometry, is presented in Section \ref{multi}, with the key results summarized in Section \ref{conclusions}.}

\section{Member selection}
\label{Targets}

\subsection{\textcolor{black}{ Motivation and basic principles}}
The key principles for selecting cluster members, outlined by \cite{1921LicOB..10..110T} over a century ago, remain relevant. The contributing factors include angular distance from the cluster center, proper motion, and position in the color–magnitude diagram. While radial velocities can be utilized, they are less reliable. Significant advancement lies in the availability of precise parallax measurements, which add another dimension to the parameter space under consideration. A combination of large proper motion, high galactic latitude, and low interstellar reddening helps in selecting Pleiades members \citep{2019A&A...628A..66L}, which are well-separated from field stars compared to other clusters \citep{2021ApJS..257...46G}.  Therefore, the approach described below would likely be less fruitful for other clusters. For a few stars, membership status remains doubtful. Due to their low number, they have been kept for further analysis.

The Gaia mission \citep{2016A&A...595A...1G} has had a tremendous impact on the census of open star clusters \citep{2020A&A...633A..99C, 2022Univ....8..111C}. Numerous papers have been dedicated to the identification of clusters and their members using various algorithms. Studies based on the Gaia EDR3 \citep{2021A&A...649A...1G} and DR3 \citep{2023A&A...674A...1G} include \cite{2022A&A...659A..59T, 2022ApJS..262....7H, 2023A&A...673A.114H, 2023A&A...675A..68V, 2023ApJS..268...30L}. However, lists based solely on Gaia data are incomplete, as evidenced by the absence of Alcyone, the brightest Pleiades star, from these lists. To address this, \cite{2023A&A...678A..75Z} additionally incorporated astrometric data from Hipparcos \citep{2007A&A...474..653V}. Yet, the problem is not confined to bright objects.  Stellar multiplicity can compromise Gaia's astrometric solutions \citep{2020MNRAS.495..321P}, leading to the exclusion of some cluster members. Therefore, a less strict approach that retains problematic entries for analysis is needed.

The Pleiades have a relatively compact core, with its 10 brightest members\footnote{Star \textit{41 Tau} ($G \sim 5.14^{\rm mag}$), positioned $5.6^\circ$ from the center, is a former Pleiades member according to \cite{2022ApJ...926..132H}, while at least \cite{2023ApJS..268...30L} and \cite{2023A&A...678A..75Z} include it in the bona fide member list.} placed inside a 1 degree radius ($1^\circ$ corresponds to 2.36 pc at the Pleiades' $d \sim 135$ pc distance). The majority of stars are within 3 pc \citep{2023A&A...673A.114H} of the cluster's center. The core is surrounded by an elongated tidal tail structure almost parallel to the Milky Way plane, with stars gradually leaving the cluster through Lagrangian points \citep{2022ApJ...931..156P}. The tidal radius, where the surrounding field density becomes less than that of the cluster, is estimated to be around 5 degrees \citep{2021A&A...645A..84M,2023A&A...678A..75Z}. Even further, a halo of unbound stars exists \citep{2022ApJ...926..132H}. Determining the membership status of these remote objects is challenging, even with accurate astrometric data. Many binary stars in the outer part of the cluster may be misclassified due to insecure astrometric solutions. Therefore, this survey focuses on the central part of the cluster within a two degree radius, corresponding to 4.7 pc at the Pleiades' distance, effectively covering the half-mass radius. This enables an analysis even when Gaia's data are not particularly reliable.

\subsection{Gaia completeness}

First, the completeness of the Gaia main source catalog is verified. It is expected to be complete within the $3 < G < 15$ apparent magnitude range, except for the secondary components in close binary systems (angular separation $\rho \lesssim 2$ arcsec) and the crowded central parts of globular clusters \citep{2021A&A...649A...5F}. The PPMXL catalog of positions and proper motions \citep{2010AJ....139.2440R}, created before Gaia, is used for validation. It includes near-infrared $J$-band photometry from 2MASS \citep{2006AJ....131.1163S}. According to the 100-Myr isochrone \citep{2012MNRAS.427..127B, 2022A&A...665A.126N}, $G-J>2.5^{\rm mag}$ is expected for Pleiades main-sequence members with $G \sim 15^{\rm mag}$. Therefore, a cutoff of $J<13^{\rm mag}$ ensures the desired $G<15^{\rm mag}$ completeness. 

In a two degree field centered on the Pleiades, 4575 PPMXL entries are found. After a nearest neighbor search within $2^{\prime\prime}$, 46 sources remain unmatched, all with proper motion $|\mu|>80$ mas yr$^{-1}$ according to PPMXL, exceeding the expected $|\mu^c| \sim 50$ mas yr$^{-1}$ for Pleiades members. To validate the cross-identification among the 4529 matched objects, the 6 outliers with $G-J>4^{\rm mag}$ are selected since a faint Gaia counterpart indicates a likely wrong identification. Indeed, for these objects, a brighter source in Gaia DR3 is present within 5 arcsec, explaining the $G-J$ inconsistency. Matches are found for all PPMXL stars within the required parameter space, confirming Gaia DR3 completeness for $G<15^{\rm mag}$ stars, except the secondary components in close binary systems. The main source Gaia DR3 catalog is further treated as the base for the sample selection. Its astrometric data processing algorithms generally do not account for stellar multiplicity \citep{2021A&A...649A...2L}. For a small fraction of sources, dedicated non-single-star solutions are available \citep{2023A&A...674A..34G}, however they largely do not affect the sample selection.

Initially, all Gaia DR3 sources within a two degree radius around the reference equatorial coordinates (Table \ref{tab:overview}) from \cite{2022A&A...659A..59T} are selected, yielding 122,711 entries.  Next, a magnitude-limited sample with $G<15^{\rm mag}$ is defined, including 6130 sources. Of these, 6047 (98.6\%) have parallax and proper motion values available in Gaia DR3. A fallback two-parameter solution without proper motion and parallax is provided when the full five- or six-parameter solution does not converge well. The reliability of full solutions can be assessed using the parameter RUWE (renormalized unit weight error), which quantifies the deviation from the standard single-star model of stellar motion. A RUWE value of $1.4$ is commonly suggested as a threshold for reliable solutions \citep{2021A&A...649A...5F}, and its excess typically marks the photocenter motion due to the presence of a companion \citep{2021ApJ...907L..33S}.

\begin{table}[h!]
\centering
\caption{Mean equatorial coordinates, proper motion (PM), and parallax of the Pleiades from the literature. \textcolor{black}{ Note that the difference is within 0.11 mas/yr for proper motion and 0.04 mas for parallax, which are well within the reported dispersion.}}
  \resizebox{\textwidth}{!}{\begin{tabular}{|l|l|l|l|l|l|}
\hline

 Parameter & \cite{2021MNRAS.504..356D} & \cite{2022ApJS..262....7H} & \Hunt 
 & \cite{2023ApJS..268...30L} & \Tarricq 
 \\ 
  \hline
 Eq. coordinates ($\alpha$ , $\delta$), $^\circ$ & 56.65 , 24.10 & 56.61 , 24.15 & 56.68 , 24.11 &56.60 , 24.08& 56.74 , 24.09\\ \hline

PM ($\mu_\alpha^c\cdot \cos\delta$), mas yr$^{-1}$ &
19.95 $\pm$ 1.11 & 20.00 $\pm$ 1.25 & 19.96 $\pm$ 1.00 $\pm$ 0.03 &19.96 $\pm$ 1.31 & 19.96 $\pm$ 0.76    \\ \hline

PM ($\mu_\delta^c$), mas yr$^{-1}$& -45.51 $\pm$ 1.18
& -45.43 $\pm$ 1.55 & -45.46 $\pm$ 1.14 $\pm$ 0.04 &-45.47 $\pm$ 1.81& -45.40 $\pm$ 0.85
\\
\hline

Parallax ($\varpi^c \pm \sigma_c$), mas & 7.35 $\pm$ 0.18 & 7.37 $\pm$ 0.21 &  7.38 $\pm$ 0.17 &7.39 $\pm$ 0.33& 7.35 $\pm$ 0.26 \\
\hline

\end{tabular}}
\label{tab:overview}
\end{table}

\subsection{Proper motion $\mu$}
\label{proper}

Measurements of relative stellar proper motion can be used to validate membership in open clusters \citep{1958AJ.....63..387V}. The Pleiades have a prominent proper motion ($\mu_\alpha^c \cos\delta \sim 20.0$ \footnote{Further on the $\cos\delta$ factor is omitted for proper motion along right ascension $\mu_\alpha$}, $\mu_\delta^c \sim -45.4$ mas yr$^{-1}$, Table \ref{tab:overview}), which is distinct from the median value for the entire $G<15^{\rm mag}$ sample ($\mu_\alpha=2.9$, $\mu_\delta=-5.5$ mas yr$^{-1}$), and allows to distinguish cluster members from randomly projected field objects very effectively. The intrinsic internal spread of velocities for Pleiades stars, while its estimated value varies in the literature, exceeds the reported uncertainties in Gaia. Indeed, the median proper motion errors of individual entries in the $G<15^{\rm mag}$ sample are 0.028 and 0.018 mas yr$^{-1}$, for $\mu_\alpha$ and $\mu_\delta$ respectively (1 mas yr$^{-1}$ corresponds to 0.64 km s$^{-1}$ at the Pleiades distance). The maximum reported errors are 0.84 and 0.59 mas yr$^{-1}$. \cite{2021ApJ...921..117T} combined a dedicated long-term radial velocity survey with astrometric data from Gaia EDR3 to conclude that the three-dimensional velocity dispersion in the Pleiades is isotropic and equals 0.83 $\pm 0.03$ km s$^{-1}$. \cite{2021A&A...645A..84M}, based on Gaia DR2 \citep{2018A&A...616A...1G} data, obtained the value of 1.4 km s$^{-1}$. \cite{2020AstBu..75..407D} adopt 2.36 km s$^{-1}$ as a velocity cutoff for cluster members, while \cite{2022ApJ...926..132H} use the same value as a threshold for currently unbound former members and suggest that the relative speed of escaped stars may reach up to 3.4 km s$^{-1}$. Both velocity dispersion and reported errors in Gaia are significantly smaller than the value of tangential speed ($\sim 32$ km s$^{-1}$), even for the most problematic astrometric solutions; therefore, proper motion appears to be the most powerful indicator of Pleiades membership.

\begin{figure}[h] 
\includegraphics[width=0.5\textwidth]{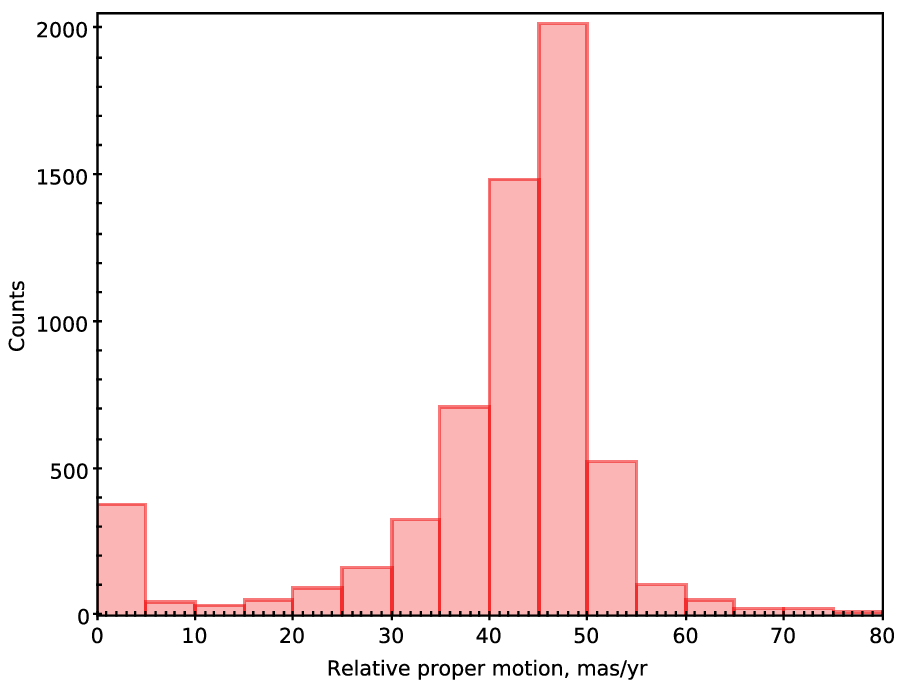}
\includegraphics[width=0.5\textwidth]{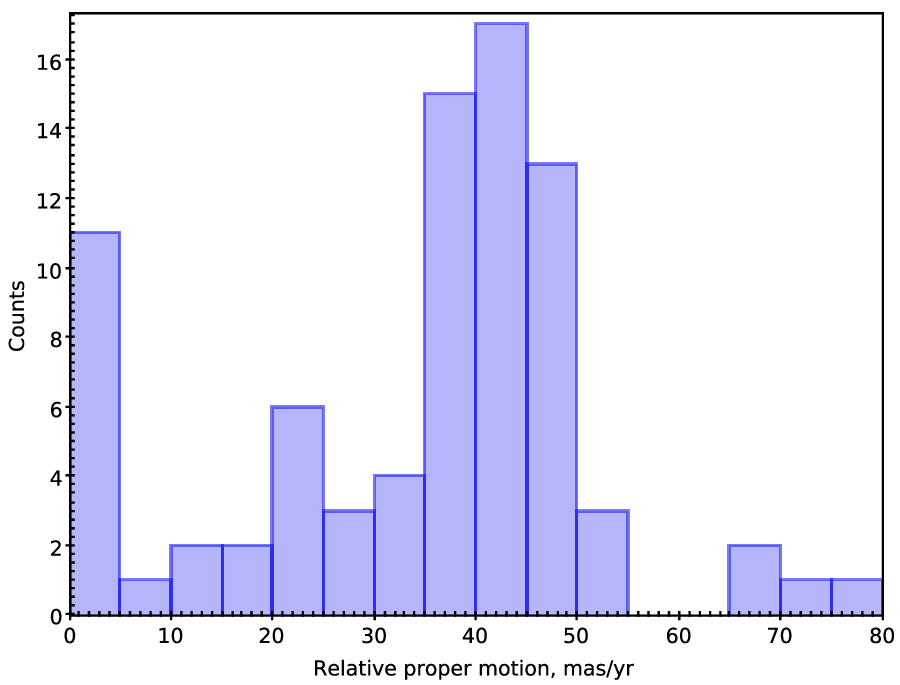}
\caption{The relative proper motion distribution ($\delta\mu$, Equation \ref{eq:pm}) for all sources with $G<15^{\rm mag}$ within a two degree radius centered on the reference coordinates from \cite{2022A&A...659A..59T}. $\delta \mu = 10$ mas yr$^{-1}$ corresponds to a tangential speed of 6.4 km s$^{-1}$ at the Pleiades distance. \textbf{Left}: 6047 stars with a full astrometric Gaia DR3 solution. \textbf{Right}: 82 objects with a two-parameter solution in Gaia DR3, for which the proper motion from PPMXL is used. Note the larger fraction of sources meeting the $\delta\mu < 5$ mas yr$^{-1}$ threshold, indicating likely cluster membership, for these entries.}
\label{fig:pm1} 
\end{figure}

\begin{equation}
\label{eq:pm}
    \delta\mu = \sqrt{(\mu_{\alpha}-\mu_{\alpha}^c)^2+(\mu_{\delta}-\mu_{\delta}^c)^2} \ {\rm ,} \ \mu_\alpha^c \sim 19.96 \ {\rm ,} \ \mu_\delta^c \sim -45.40 \ {\rm mas \ yr^{-1}}
\end{equation}

The value of Equation \ref{eq:pm} is calculated to define the proper motion of each source, $\mu_{\alpha,\delta}$, relative to the reference values $\mu_{\alpha,\delta}^c$ from \cite{2022A&A...659A..59T}. The resulting $\delta\mu$ histogram clearly shows a bimodal distribution (Figure \ref{fig:pm1}). Of the 6047 stars, 377 have $\delta\mu < 5$, and 419 meet the $\delta\mu < 10$ mas yr$^{-1}$ threshold according to Gaia DR3 astrometric data. \textcolor{black}{ Some of the randomly projected objects may have proper motion similar to cluster members. For a rough estimate of the corresponding contamination from field stars, the reference proper motion is deliberately changed to the offset values $\mu_\alpha^c=10$ and $\mu_\delta^c=-35$ mas yr$^{-1}$.} In this case, the number of entries with $\delta\mu < 5$ mas yr$^{-1}$ drops to 10 objects. Since the actual proper motion of the Pleiades is larger, and the difference with field stars is even more pronounced, the expected contamination level in the $\delta \mu<5$ mas yr$^{-1}$ subsample should not exceed a few entries.

A large tangential velocity enables the use of long-term ground-based observations for objects with problematic Gaia solutions. A detailed analysis of the PPMXL catalog \citep{2010AJ....139.2440R}, which contains the required data, and its comparison \citep{2019AJ....157..222S} with Gaia, exceeds the scope of this study. PPMXL has a lower angular resolution than Gaia. Since the closest neighbor in a 2 arcsec radius is selected around objects of the $G<15^{\rm mag}$ sample, typically one unresolved PPMXL source matches both stars of the resolved pair in Gaia. Overall, counterparts are found for 6029 out of 6130 (98.4\%) entries. The PPMXL catalog is normally used when Gaia solutions lack proper motion and parallax. Particularly, for 11 stars with two-parameter solutions $\delta\mu < 5$ mas yr$^{-1}$ is met, if $\mu$ from PPMXL is introduced (Figure \ref{fig:pm1}). In some cases, PPMXL data suggest cluster membership despite an unconvincing Gaia solution. Thus, V1228 Tau has a $\delta \mu=16.6$ mas yr$^{-1}$ according to Gaia DR3, indicating a relative velocity above 10 km s$^{-1}$, but a large RUWE of 5.7 questions this. PPMXL data reduce $\delta \mu$ to 1.5 mas yr$^{-1}$, a plausible value for a Pleiades star.

\subsection{Parallax $\varpi$}

\begin{figure}[h] 
\centering
\includegraphics[width=.49\textwidth]{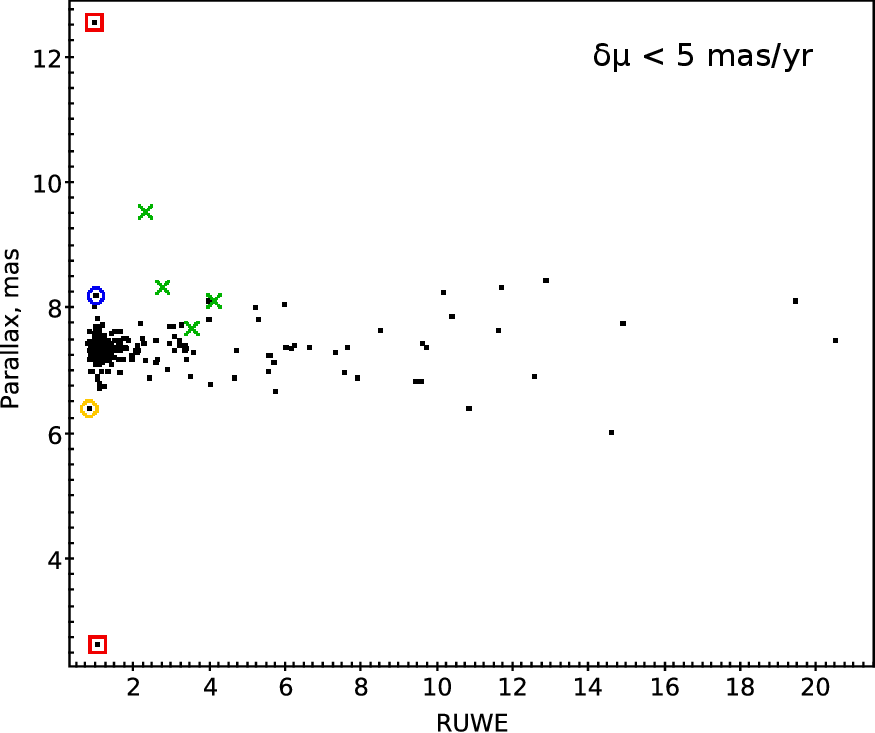} 
\includegraphics[width=.49\textwidth]{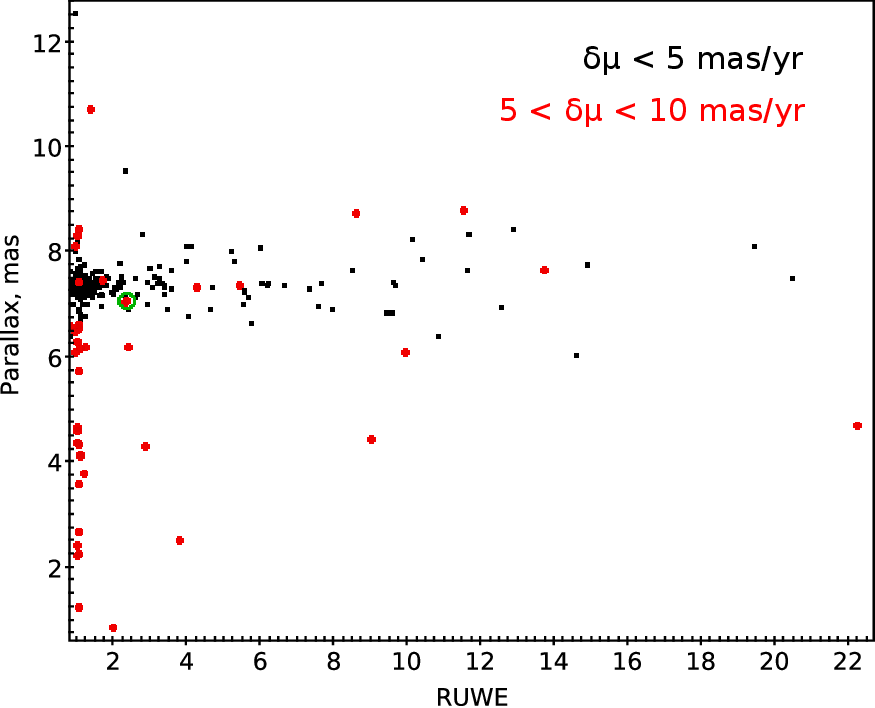}
\caption{Parallaxes and RUWE for the $G<15^{\rm mag}$ sample, restricted by relative proper motion. \textbf{Left}: $\delta\mu <5$ mas yr$^{-1}$ (Equation \ref{eq:pm}). 377 stars with available parallax in Gaia DR3 are shown. Two entries with reliable low-RUWE solutions, $\varpi=12.55\pm0.03$ and $\varpi=2.65\pm0.02$ mas, are marked by \textcolor{red}{$\square$} and considered to be foreground and background objects, respectively. Their Gaia DR3 radial velocities, $\nu=-6.1\pm0.2$ and $27.2 \pm 1.9$ km s$^{-1}$, along with positions in the color-magnitude diagram (Figure \ref{fig:cmd}), confirm these sources are unrelated to the Pleiades. Among entries with $\rm{RUWE} < 2.5$, the star marked by \textcolor{blue}{$\bigcirc$} shows the largest parallax, $\varpi=8.19 \pm 0.02$ mas, which places it less than 14 pc from the cluster's center. Since its radial velocity perfectly matches that of the cluster, it is considered to be a probable member. The smallest $\varpi$ for the star with low RUWE is $6.40 \pm 0.02$ mas, corresponding to a distance of 20 pc from the center. Its Gaia DR3 radial velocity $\nu=-0.5\pm0.3$ also suggests that this star (marked by \textcolor{orange}{$\bigcirc$}) is unrelated to the Pleiades. Finally, the brightest ($G<4.5^{\rm mag}$) cluster members are marked by \textcolor{green}{$\times$}, all of which show RUWE excess. Two stars with $G<4.5^{\rm mag}$ are not shown: Alcyone, which has a two-parameter solution, and Asterope, with $\delta \mu \sim 7.5$ mas yr$^{-1}$ according to Gaia DR3. \textbf{Right:} Previously discussed sources with $\delta\mu <5$ mas yr$^{-1}$ are shown with black dots, objects with $5<\delta\mu <10$ mas yr$^{-1}$ are colored \textcolor{red}{\bf red}. While the newly introduced subsample contains many background sources, several stars such as  Asterope (marked \textcolor{green}{$\bigcirc$}, $G=4.2^{\rm mag}$) are certain Pleiades members.}
\label{fig:pm5} 
\end{figure}

While a small $\delta \mu $ value strongly indicates Pleiades membership, other parameters should be considered for further validation. As with proper motion, different authors agree on the mean Pleiades parallax $\varpi^c \sim 7.35 \ \rm{mas,}$ while its reported dispersion $\sigma_\varpi$ is more model-dependent (Table \ref{tab:overview}). Considering the tidal radius estimate of 11.3 -- 12.5 pc, depending on unresolved binary stars' contribution \citep{2023A&A...678A..75Z}, parallaxes of bona fide members are expected within $6.75 \lesssim \varpi \lesssim 8.15$ mas, distinctly different from the median value $\varpi=1.20$ mas for the entire $G<15^{\rm mag}$ sample.  However, open clusters lack a sharp boundary, and considering parallax uncertainties in Gaia, this range can extend further, especially for problematic solutions with large RUWE. Simulations by \cite{2022MNRAS.513.2437P} show that unresolved components may influence parallax and, to a lesser extent, proper motion estimation through the standard single-star Gaia pipeline, potentially placing the real value of $\varpi$ far beyond the reported uncertainties. Among the 377 Gaia DR3 sources with $\delta\mu<5$ mas yr$^{-1}$ shown in Figure \ref{fig:pm5}, only two entries distinctly stand out with reliable low-RUWE solutions and outlying parallaxes, considered unrelated to the Pleiades, while the rest concentrate around $\varpi^c$ as expected.  Expanding to the $5<\delta\mu<10$ mas yr$^{-1}$ domain significantly increases the fraction of background stars with small parallaxes; nevertheless, several sources clearly fit the pattern of probable cluster members.

\subsection{Radial velocity $\nu$}

Analysis of heliocentric radial velocities $\nu$ further constrains the sample, but their interpretation is complex due to inhomogeneous data.  Several factors hinder the correct determination of $\nu$. Hot stars of spectral types B and A have few spectral lines and possess large rotational broadening \citep{2020ApJ...901...91T, 2023A&A...674A...7B}. At the faint end of the $G<15^{\rm mag}$ sample, the factor of low signal-to-noise ratio dominates. The presence of a bright neighbor within a few arcseconds may cause the spurious results \citep{2019MNRAS.486.2618B}. Stellar activity causes intrinsic radial velocity wobble \citep{2020PASJ...72..104T}. Finally, binary and multiple stars show variation of $\nu$ due to the motion of components around a common center of mass. Thus, some reported radial velocities can be misleading and should be treated with caution.

While one catalog does not have adequate completeness for the entire sample, combining data sets allows achieving it. Radial velocities from Gaia DR3 \citep{2023A&A...674A...5K} are available for 5365 objects (87.5\% fraction). Reported errors depend heavily on the apparent magnitude: the median value is within 1 km s$^{-1}$ for $G<12^{\rm mag}$ sources, but exceeds 5 km s$^{-1}$ for the $14<G<15$ magnitude range. Data reliability is particularly questionable for young, active low-mass stars. A comparison by \cite{2022MNRAS.517.1946K} shows that the difference between Gaia DR3 and APOGEE can exceed 10 standard errors or 10 km s$^{-1}$ in absolute value for stars presumed to be single Pleiades members. 

An important benchmark area, the Pleiades, has been extensively covered by spectroscopic surveys.  The Large sky Area Multi-Object fiber Spectroscopic Telescope (LAMOST\footnote{The latest to date publicly available Data release 9 version 2.0 is used for LRS and MRS template estimates (\textit{rv\_b1}, \textit{rv\_r1}, \textit{rv\_br1}), while version 1.1 is used to obtain \textit{rv\_lasp1}.}, \cite{2012RAA....12..723Z})  provides several data products containing radial velocities. $\nu$ obtained through the LAMOST stellar pipeline (LASP, \cite{2014IAUS..306..340W}) during the low-resolution ($\lambda/ \delta \lambda \sim 1800$) survey is available for 74\% of sample objects with $G<15^{\rm mag}$. Additionally, four different estimates of radial velocity \citep{2019ApJS..244...27W}, obtained either through the cross-correlation method with synthetic spectra in different wavelengths (\textit{rv\_b1}, \textit{rv\_r1}, and \textit{rv\_br1}) or by the LASP (\textit{rv\_lasp1}) from the medium-resolution ($\lambda/ \delta \lambda \sim 7500$) survey (MRS, \cite{2020arXiv200507210L}), have higher accuracy at the cost of lower completeness. LAMOST provides individual epoch measurements. For cluster membership classification, the median value is used. The native cross-identification with Gaia DR3 appears to be incomplete, and the following routine is used. LAMOST counterparts are searched according to coordinates in columns \textit{ra, dec} within 3 arcsec from Gaia sources. Due to lower angular resolution, one LAMOST source may correspond with two Gaia entries. In such cases, the pair with the lower matching distance is selected. Although this algorithm may produce a few incorrect or missing matches, since radial velocity analysis is largely collateral for this study, and LAMOST is not the sole data set, the described approach is maintained.

Sloan Digital Sky Surveys Data Release 17\footnote{File \textit{allStar-dr17-synspec\_rev1.fits} which includes built-in identification with Gaia is used.} \citep{2022ApJS..259...35A} includes observations from The Apache Point Observatory Galactic Evolution Experiment  (APOGEE; \cite{2017AJ....154...94M}),  which targeted the Pleiades in several campaigns \citep{2014ApJ...794..125C, 2019AJ....157..196K, 2020AJ....159..199D, 2022AJ....164...85M, 2022MNRAS.517.1970S, 2023AJ....165...51R}. While its general completeness is lower compared to Gaia or LAMOST, the coverage for probable members almost reaches Gaia's level, providing higher accuracy. For a fairly low fraction of stars, radial velocities are available in GALAH+ Third data release \citep{2021MNRAS.506..150B}. Either template measurements based on stacked observed GALAH spectra \textit{rv\_nogr\_obst} or those obtained through synthetic spectra comparison \textit{rv\_sme\_v2} are used.

\begin{figure}[h] 
\centering
\includegraphics[width=.7\textwidth]{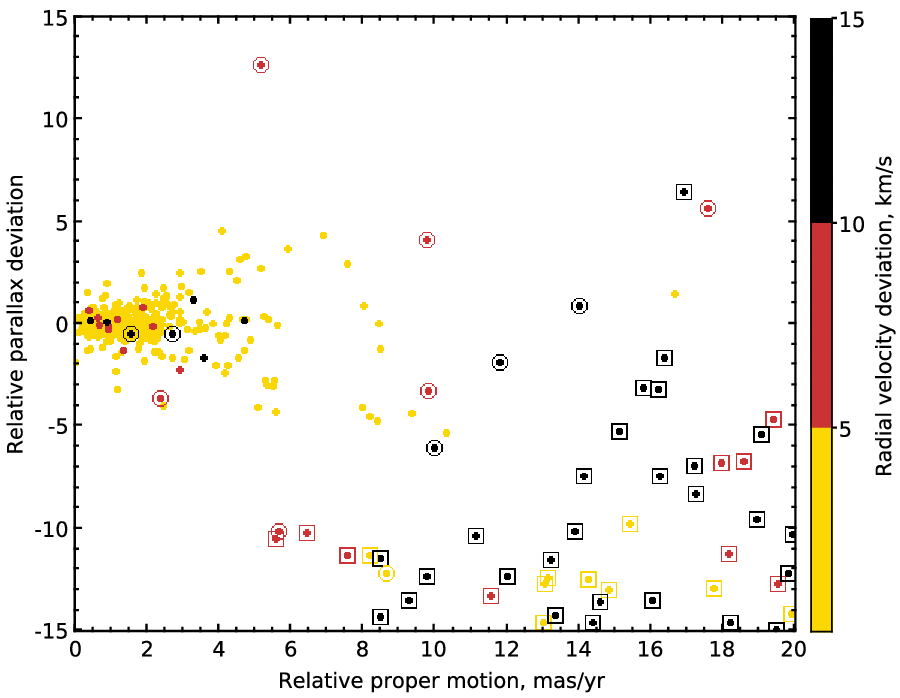}
\caption{Deviation of radial velocity $\delta \nu = |\nu-\nu^c|$ (color-coded) from the reference $\nu^c=5.69$ km s$^{-1}$, depending on the relative proper motion $\delta \mu$ (Equation \ref{eq:pm}), and parallax deviation, defined as $\Delta \varpi / \sigma_\varpi = (\varpi-\varpi^c) / \sqrt{\sigma^2 + \sigma_c^2}$. $\varpi^c$ and $\sigma_c$ are the cluster's average parallax and dispersion according to \cite{2022A&A...659A..59T} (Table \ref{tab:overview}), $\varpi \pm \sigma$ is the reported Gaia DR3 parallax for a given star. Sources with $\delta \mu$ and $\Delta \varpi / \sigma_\varpi$ values close to zero are superimposed in the plot. The absolute majority of them fall into the $\delta \nu<5$ km s$^{-1}$ range, but a few stars with deviating radial velocities are still present. For most of them, it can be explained by the disturbance from the neighboring source within 1 -- 2 arcsec; however, two stars marked by $\bigcirc$ show $\delta \nu \gtrsim 15$ km s$^{-1}$ according to both Gaia and APOGEE data, suggesting they may be field stars. Their faint magnitudes ($G \sim 14.7$ and $14.8^{\rm mag}$) imply large uncertainties, while their position in the color-magnitude diagram (Figure \ref{fig:cmd}) is consistent with being a cluster member, so their membership is not completely ruled out. In the plot, bare points $\cdot$ are used for probable cluster members. Sources considered to be field stars are marked with $\square$, while stars surrounded by $\bigcirc$ have controversial status. Notably, all probable members meet the $|\Delta \varpi / \sigma_\varpi |< 5.5$ condition, regardless of the $\delta \mu$ value. Several objects outside this range are given controversial status if their radial velocity and position in the color-magnitude diagram do not contradict cluster membership.}
\label{fig:target} 
\end{figure}

Finally, \cite{2021ApJ...921..117T} provides results from a four-decade observational campaign conducted by the Center for Astrophysics (CfA) and CORAVEL \citep{1979VA.....23..279B}. This extensive duration allowed the discovery of long-period spectroscopic binaries missed by other observers. The obtained mean radial velocity, $\nu^c = 5.69 \pm 0.07$ km s$^{-1}$, serves as a reference throughout this paper, \textcolor{black}{ matching estimates based on Gaia and APOGEE data, $5.34 \pm 1.10$ and $6.1 \pm 0.9$ km s$^{-1}$ \citep{2024A&A...686A..42H, 2020AJ....159..199D}.} Unlike proper motion or parallax, this value closely matches the median for the entire $G<15^{\rm mag}$ sample, $\nu = 3.0$ km s$^{-1}$, limiting the role of $\nu$ in constraining cluster membership. 

Overall, ten measures (five estimates from LAMOST, two from GALAH+, and one coming from Gaia DR3, APOGEE and joint CfA \& CORAVEL survey) of radial velocity are considered. The coverage for probable members with consistent proper motion is essentially complete, with the exception of bright sources with $G<6.5^{\rm mag}$. Most stars have measurements from two or more surveys. The value closest to the reference $\nu^c = 5.69$ km s$^{-1}$ is selected. Although this approach biases the $\nu$ distribution, it reduces the chance of erroneous exclusions from the cluster member list. 

Previous analysis (Figure \ref{fig:pm5}) showed that sources with low relative proper motion generally have parallax close to the cluster's average, suggesting Pleiades membership. The introduction of $\nu$ refines the distinction between members and field objects (Figure \ref{fig:target}). Most stars with astrometric solutions consistent with cluster membership have radial velocities near the reference value $\nu^c$. A few exceptions occur among stars with $\delta \mu <5$ mas yr$^{-1}$; however, some spurious measurements were anticipated. Some of these are related to spectroscopic binary stars, for which an accurate determination of the system's $\nu$ is hardly possible without rarely available long-term observations. Approximately within the $\delta \mu <10$ mas yr$^{-1}$ range, sources with parallax close to $\varpi^c$ largely show radial velocities expected for Pleiades members. Beyond this range, objects with large radial velocity deviation $\delta \nu = |\nu-\nu^c|$ dominate.  For stars in the boundary region, consideration of their positions in the color-magnitude diagram (Section \ref{cmd}) is desirable.

\subsection{Color-magnitude diagram}
\label{cmd}
\begin{figure}[ht] 
\centering
\includegraphics[width= .7 \textwidth]{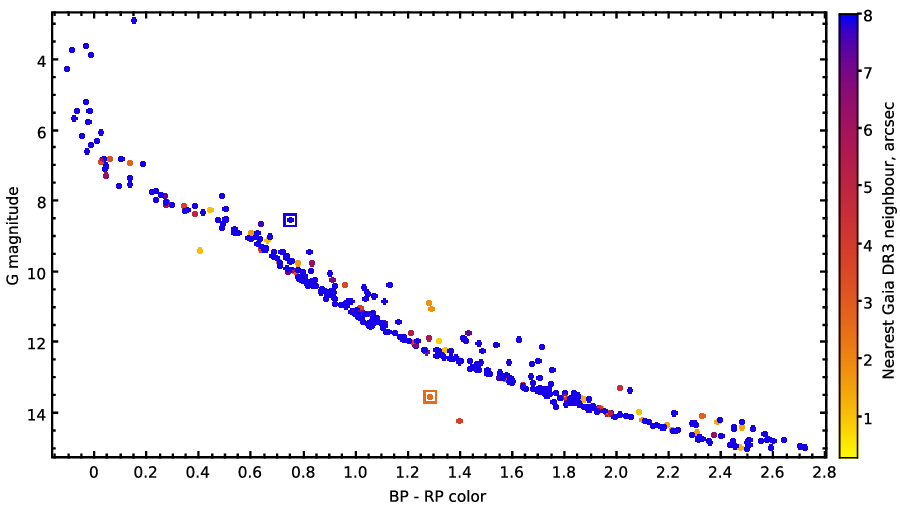}
\caption{Color-magnitude diagram for stars with $\delta \mu < 5$ mas yr$^{-1}$ (Equation \ref{eq:pm}). \textcolor{black}{ PPMXL data are used for the two-parameter Gaia DR3 solutions.} Two entries marked by $\square$ are unrelated to the Pleiades based on their parallaxes and radial velocities. The presence of a close neighbor in Gaia, potentially creating an artifact in $c=B_P-R_P$ estimation, is color-coded (Section \ref{cmd}).} 
\label{fig:cmd} 
\end{figure}

Since the Pleiades represent a simple (single) stellar population \footnote{Single stellar population includes binary and multiple systems in this sense.} \citep{2012A&ARv..20...50G}, united in its age and chemical abundance, the constituent stars naturally form a distinct sequence in the color-magnitude diagram (Figure \ref{fig:cmd}), which is a useful tool for diagnostics of cluster parameters \citep{2016csss.confE.102B}. The observed relation has a sizeable width for several reasons, such as the intrinsic size of the cluster\footnote{Adopting a 10 pc tidal radius and a 135 pc mean distance, the total expected scatter is $\sim 0.3^{\rm mag}$ (Section \ref{mass}).}, unresolved stellar multiplicity \citep{1991ApJ...372..487R}, the influence of diverse stellar rotation \citep{2009MNRAS.398L..11B}, or incorrect multicolor photometry. Regarding the latter, color estimates in Gaia are distorted by blending effect and should be treated carefully. $B_P$ and $R_P$ fluxes are efficiently measured through aperture photometry in a fairly large $3.5 \times 2.1^{\prime\prime} $ window \citep{2021A&A...649A...3R}, while $G$ magnitudes are obtained through line spread function fit optimized for point sources \citep{2021A&A...649A..11R}. As the result for a resolved binary with angular separation within few arcseconds,  $B_P$ and $R_P$ magnitudes represent the combined flux from a system, while an angular resolution in $G$ band allows to obtain individual brightness of a star. The resultant bias heavily affects the reported $B_P-G$ and $G-R_P$ colors, rendering them impractical for member selection. Nonetheless, the analysis of color-color diagrams is valuable for subarcsecond multiplicity diagnostics (Section \ref{multi}).

\subsection{Member list and comparison with earlier works}
\label{sample}

\begin{table}[h!]
\centering
\caption{List of surveyed Pleiades members. The ten brightest entries are shown; the full table is available at \href{https://github.com/chulkovd/Pleiades/}{Github}. Values are left blank if missing, e.g., due to a two-parameter solution. Columns 1 - 3: designation, apparent $G$ magnitude, and RUWE in Gaia DR3. Column 4: relative proper motion (Equation \ref{eq:pm}). Columns 5 and 6: parallax and radial velocity deviation (Figure \ref{fig:target}). Column 7: source of radial velocity data. Columns 8 and 9: angular separation and apparent magnitude of the nearest Gaia DR3 source within $\rho<10^{\prime\prime}$. Columns 10 - 16: cluster membership according to papers in Table \ref{tab:surveys}. Stars with membership probability below 0.5 are marked \textit{2}. Column 17 marks entries (14 overall) with controversial membership status.}
    \label{tab:sample}
 \resizebox{\textwidth}{!}{\begin{tabular}
{|c|c|c|c|c|c|c|c|c|c|c|c|c|c|c|c|c|c|}
\hline 
1&2&3&4&5&6&7&8&9&10&11&12&13&14&15&16&17
      \\  \hline
        Gaia DR3 Source &$G$&RUWE   &$\delta \mu$&$\Delta \varpi / \sigma_\varpi$ &$|\nu -\nu_c|$&data set&$\rho$&$G_n$&i&ii&iii&iv&v&vi&vii&C
        \\ \hline

66714384142368256&2.896&-&-&-&-&-&-&-&0&0&0&0&0&0&0&1 \\ \hline
66526127137440128&3.616&4.117&1.186&1.4&-&-&-&-&0&0&1&2&1&1&1&1 \\ \hline
65271996684817280&3.698&2.768&0.897&1.94&0.9&Gaia&-&-&0&0&1&2&2&1&2&1 \\ \hline
65283232316451328&3.863&3.542&0.462&0.78&0.2&CfA&-&-&1&0&1&1&1&1&1&1 \\ \hline
65205373152172032&4.173&2.33&7.512&-0.74&-&-&-&-&0&0&0&0&0&1&0&1 \\ \hline
65296907494549120&4.261&2.315&4.089&4.55&4.3&CfA&-&-&0&0&0&0&0&1&0&1 \\ \hline
66529975427235712&5.203&1.303&2.299&-0.39&-&-&-&-&1&1&1&2&1&1&1&1 \\ \hline
64940906245415808&5.428&1.071&1.735&1.2&1&CfA&-&-&1&1&1&1&1&1&1&1 \\ \hline
65287458566524928&5.441&0.963&0.446&0.11&10.8&Gaia&-&-&1&1&1&1&1&1&1&1 \\ \hline
69812945346809600&5.64&0.899&0.77&-0.49&2.1&CfA&-&-&1&1&1&1&1&1&1&1 \\ \hline

    \end{tabular}}
    
\end{table}

\begin{figure}
\centering
\includegraphics[width=0.55\textwidth]{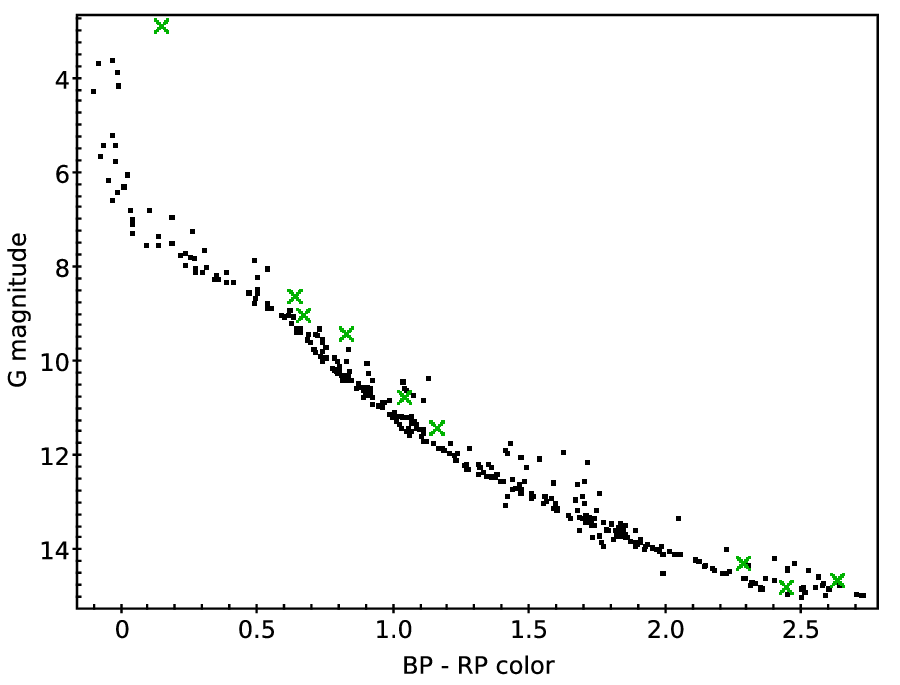}
\caption{351 isolated sources with no Gaia DR3 neighbors within 5 arcsec considered probable Pleiades members. Stars with two-parameter solutions (marked by \textcolor{green}{$\times$}) are displaced toward brighter magnitudes, as expected for unresolved binary systems.
}
\label{fig:cmd2} 
\end{figure}

In summary, the selection of the cluster members predominantly relies on astrometric data. Objects with available Gaia parallax and proper motion are chosen according to their appearance in Figure \ref{fig:target}. Particular attention is given to sources with a two-parameter solution in Gaia DR3, comprising a total of 83 entries in the initial $G<15^{\rm mag}$ sample. Of these, 23 appear within 2 arcsec of another Gaia entry. In 8 cases (4 pairs), both sources are two-parameter. However, $\mu$ from the PPMXL catalog allows for rejecting their cluster membership. For the other 15 stars, the second source in a pair has a full astrometric solution in Gaia, and 4 of them are attributed to the Pleiades along with a component. Assuming a uniform distribution of 6130 stars with $G<15^{\rm mag}$ across a two degree field, the probability of finding any optical pairs with $\rho<2^{\prime\prime}$ in the sample is less than 50\%. Isolated two-parameter sources are classified based on proper motion in PPMXL, radial velocity, and their position in the color-magnitude diagram (Figure \ref{fig:cmd2}).

\begin{table}
     \caption{    
    The number of Pleiades members compared with previous studies based on Gaia EDR3 or DR3, a two degree radius explored in this paper, and $G<15^{\rm mag}$ stars are considered. Some surveys explicitly provide the cluster's membership probability $p$. \cite{2022ApJ...926..132H} divide entries into current members and escapee-candidate categories. Bona fide members, according to \cite{2023A&A...678A..75Z}, are counted in the column $p \geq 0.5$. A detailed breakdown for each entry is available in Table \ref{tab:surveys}.}
    \label{tab:surveys}
    \centering
    
    \begin{tabular}{|c|c|c|c|}
        \hline
         &Member list& N ($p\geq0.5$)  &N ($p\leq0.5$) \\
        \hline
         i&\cite{2022ApJ...926..132H}& 348 & 2* \\
\hline
ii&\Tarricq
         & 273 & 4\\
\hline        iii&\cite{2022ApJS..262....7H}& 371 & -- \\
         \hline
iv&\Hunt&301&71\\          \hline

v&\Groeningen &326&8\\
\hline
vi& \cite{2023ApJS..268...30L} & 378& --\\
\hline
vii&\Zerjal&347&24**\\
\hline
\multicolumn{2}{|c|}{This study}&409&14\\

\hline
    \end{tabular}
   
\end{table}

A total of 409 Gaia DR3 sources (Table \ref{tab:sample}) are considered probable Pleiades members, with 14 given a controversial status. While a few selected objects may be formally unbound to the cluster, this does not impact the multiplicity analysis if they share the Pleiades' origin and have a comparable distance from the Sun. The obtained member list contains at least 31 more entries compared to previous works (Table \ref{tab:surveys}). 18 stars, deemed probable members, are not included in any of the lists mentioned above. Among them, 14 have a two-parameter solution. The remaining four stars are: V1228 Tau ($G=7.63^{\rm mag}$), 
V679 Tau ($G=11.73^{\rm mag}$), 2MASS J03431902+2226572 ($G=11.79^{\rm mag}$), and 2MASS J03402958+2333040 ($G=13.72^{\rm mag}$). The first three are known multiple or binary stars, which explains their unconvincing astrometric solutions marked by the RUWE excess. The latter star has a reasonably low RUWE value (1.25); however, its proper motion according to PPMXL ($\delta \mu = 2.7$ versus $\delta \mu = 9.4$ mas yr$^{-1}$ with Gaia DR3) and radial velocity from various sources are consistent with cluster membership. The probable members list, together with objects of controversial status, includes every entry considered in the papers referenced in Table \ref{tab:surveys}.

\section{Theoretical isochrone}
\label{isochrones}
\subsection{Isochrone selection}
The defining feature of star clusters is the common origin of their members, which provides an excellent opportunity to test stellar evolution models \citep{2022Univ....8..111C} or the quality of catalog data \citep{2023A&A...677A.162B}. While an intracluster age gradient \citep{2018MNRAS.476.1213G} and multiple star formation episodes are observed in the youngest clusters with an age of several Myr \citep{2019A&A...627A..57J}, the corresponding effect remains unnoticeable for older objects, such as the Pleiades ($t\sim 100$ Myr). Therefore, the sample is treated as a simple, equal-age population. \textcolor{black}{ Although age is notably hard to determine for a lone star \citep{2010ARA&A..48..581S}, there are a variety of age-dating methods for star clusters \citep{2016EAS....80..115B}. Among them, the lithium depletion technique appears to be promising for the Pleiades-age clusters \citep{2014EAS....65..289J}. 
Recent estimates of the Pleiades' age with this method are rather consistent: $t=112 \pm 5$ \citep{2015ApJ...813..108D}, $127^{+6}_{-10}$ \citep{2022A&A...664A..70G}, $93^{+9}_{-10}$ Myr \citep{2023MNRAS.523..802J}. Isochrone fitting \citep{1981A&A....97..235M} tends to bring larger uncertainty, but largely does not contradict $t \sim 100-125$  Myr estimate \citep{2018A&A...616A..10G, 2023A&A...677A.162B}. However, despite the Pleiades representing a benchmark for isochrone testing, the sequence for low-mass stars ($M\lesssim0.6 M_\odot$) is not well-fitted by the existing models \citep{2012MNRAS.424.3178B}.}

\begin{figure}[h] 
\includegraphics[width=.7\textwidth]{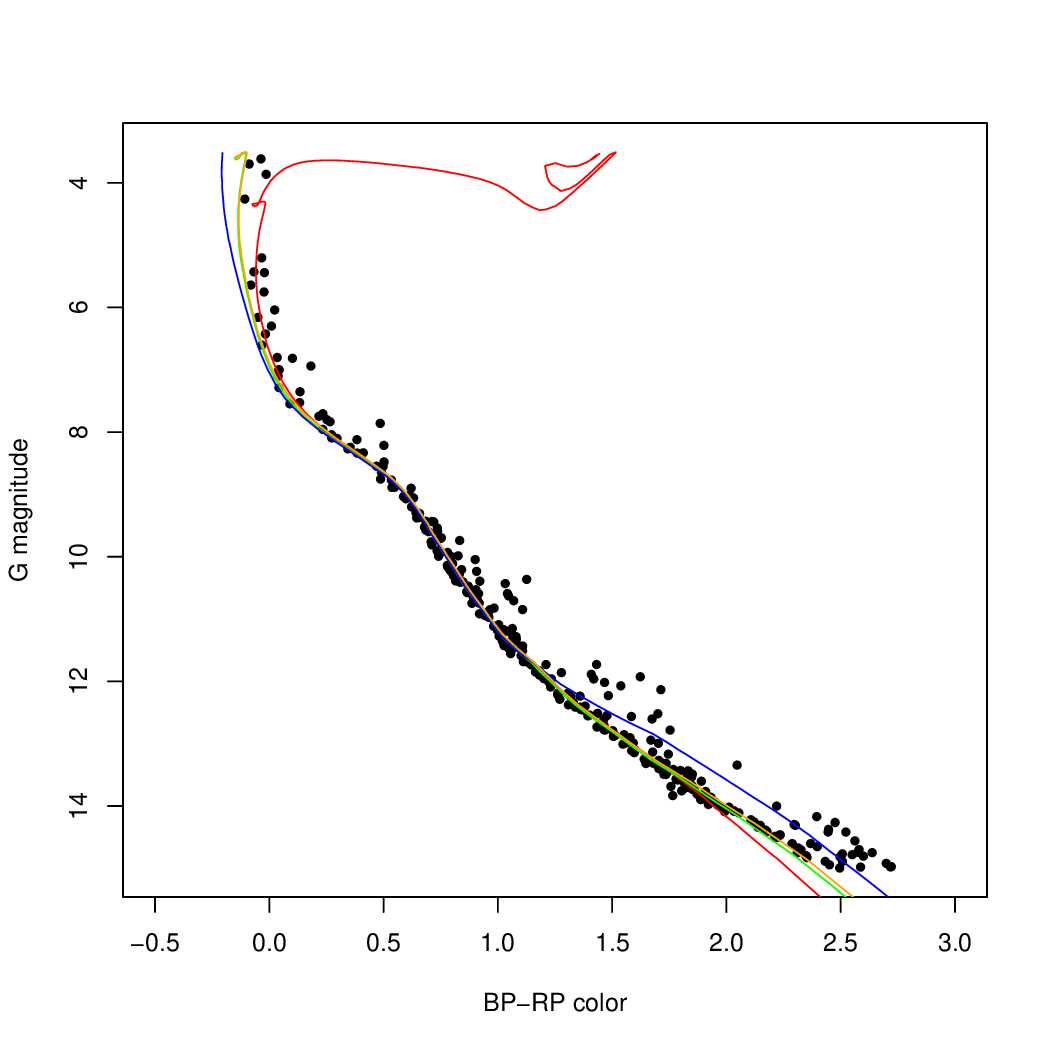}
\caption{PARSEC (version 2.0,  \cite{2012MNRAS.427..127B, 2022A&A...665A.126N}) isochrones with rotation $\Omega/\Omega_c=0.3$ (inclination $60^\circ$) for ages $t=$ \textcolor{blue}{50}, \textcolor{green} {100}, and \textcolor{red} {200} Myr at solar metallicity. The model with  \textcolor{orange}{enhanced [Fe/H]} = 0.05, $t=100$ Myr is adopted for further calculations. \textcolor{black}{ It provides reasonable agreement for sources with apparent magnitude $G>7^{\rm mag}$, thus covering the binary systems considered in Table \ref{tab:outliers}.} A distance of 135 pc ($5.65^{\rm mag}$ modulus) and an extinction of $A_V=0.15^{\rm mag}$ are applied.}
\label{fig:PARSEC} 
\end{figure}

Besides age, other parameters such as metallicity, distance, and interstellar extinction should be defined, as they alter the cluster's appearance in the color-magnitude diagram (Figure \ref{fig:PARSEC}). Open clusters are expected to be metallicity-homogeneous within [Fe/H] $<$ 0.05 \citep{2016MNRAS.457.3934L, 2018ApJ...863..179S}. 
The Pleiades show a solar or marginally enhanced metallicity according to spectroscopic observations \citep{2009AJ....138.1292S, 2022A&A...668A...4F}. A distance of 135 pc, which converts into a $5.65^{\rm mag}$ modulus, is applied to all stars. The interstellar medium around the Pleiades has a complex structure \citep{2003ApJS..148..487W}, possibly contributing to extinction $A_V$ and reddening non-uniformity within the cluster \citep{2008AJ....136.1388T}. \textcolor{black}{ This factor potentially introduces additional dispersion to the observed magnitudes on a scale of a few hundredths of a magnitude for sources in different parts of the cluster but should not affect the relative flux of binary star components.}
Following \cite{2020MNRAS.499.1874M}, $A_V=0.15^{\rm mag}$ is adopted. Extinction in different wavelengths is then calculated according to \cite{1989ApJ...345..245C} and \cite{1994ApJ...422..158O} through the CMD\footnote{http://stev.oapd.inaf.it/cgi-bin/cmd} interface.

Fortunately, the choice of a particular model has a limited impact on mass calculation. As shown in \cite{2023A&A...678A..75Z}, the $G-R_p$ color-mass relation for ages between 50 Myr and 1 Gyr shows a small enough variation for a rough mass estimation. Though several theoretical models are proposed in literature, their detailed comparison and underlying physics analysis are beyond this paper's scope. After visual inspection, the PARSEC \citep{2012MNRAS.427..127B, 2022A&A...665A.126N} $t=100$ Myr isochrone is chosen for further evaluation (Figure \ref{fig:PARSEC}). Recent work by \cite{2024MNRAS.528.4272H} agrees with the preference for the PARSEC model over MIST \citep{2016ApJ...823..102C} in the low-mass regime. Given the presence of unresolved binary stars in the sample, which show higher brightness and redder color compared to single stars, preference is given to the lower bound of the observed sequence during fitting. The adoption of a slightly enhanced metallicity [Fe/H]$=0.05$ improves the agreement for low-mass stars. Stellar rotation is regarded as one of the reasons for sources displacement in the color-magnitude diagram \citep{2014AJ....148...30K}. The PARSEC model accounts for this; however, the estimated impact of moderate spin with $\Omega/\Omega_c=0.3$ is within $0.02^{\rm mag}$ in comparison to non-rotating stars for $M<4M_\odot$, where $\Omega_c$ is the breakup angular velocity. An inclination of $60^\circ$ is adopted for calculations, which is the average angle for an isotropic distribution \citep{2021ApJ...923...23H}. Generally, the choice of a theoretical isochrone is somewhat arbitrary, and models with slightly different parameters can be justified as well.

\subsection{Mass estimation}
\label{mass}

After selecting the isochrone, a mock population of single and binary stars is generated.  Since some of the considered sources have poorly defined or missing parallaxes, a uniform distance of $d=135$ pc is applied. Assuming a radius of $r \sim 10$ pc, an object at the cluster's near edge appears $5\log_{10}\frac{d+r}{d} \sim 0.15^{\rm mag}$ brighter than an identical star near the center; this value is taken as the uncertainty of the $G$. Monte Carlo sampling with error margins of quantiles 0.15 and 0.85 is utilized to accommodate uncertainties. Using the chosen theoretical isochrone (Figure \ref{fig:PARSEC}), apparent magnitudes in various passbands are computed across a dense mass grid, linear interpolation is applied when necessary. The $G$ magnitude thus directly constrains the mass of isolated stars. Subsequently, the mass ratio of components $q=m_2/m_1$, $0<q<1$, is introduced.  Depending on the angular separation and possibly other factors, the reported Gaia magnitude may correspond to individual component flux or the total brightness of the binary system. This distinction can be discerned from the source's position in the color–magnitude diagram (Figure \ref{fig:cmd}) if the flux from the secondary star is large enough. Primary, secondary, and combined magnitudes are computed across a two-dimensional grid of $m_1$ and $q$. For passbands other than $G$, the relative magnitude contrast between the primary and secondary components is calculated. The values of $m_1$ and $q$ that agree with observed magnitude and contrast are selected. This methodology is validated on V1282 Tau, a spectroscopic binary with high-quality multicolor contrast data and an independent mass ratio estimate from orbital solutions by \cite{2020ApJ...898....2T}. The adopted magnitude cutoff of $G<15^{\rm mag}$ corresponds to masses $m \gtrsim 0.47 M_\odot$ for single stars or $m_1 \gtrsim 0.38 M_\odot$ in the case of unresolved twin binaries.

\section{Multiplicity analysis}
\label{multi}

As discussed in Section \ref{cmd}, the Pleiades stars form a distinct relation in the color-magnitude diagram, and an offset from the main sequence is regarded as a marker of peculiarity, such as stellar multiplicity, incorrect photometric data, or a wrong assessment of membership status. The related analysis is an essential part of numerous papers \citep{2020ApJ...903...93N, 2021AJ....162..264J, 2022MNRAS.516.5637E, 2023A&A...672A..29C, 2023ApJS..268...30L, 2023MNRAS.525.2315A} concerning the binary population of open clusters. Remarkably, at least \cite{2023A&A...675A..89D} and \cite{2023AJ....166..110P} emphasize the often-neglected contribution  of resolved binary stars in nearby clusters. While the mentioned works focus on Gaia photometric data, \cite{2023AJ....165...45M} introduce a multicolor approach. However, the common assumption that 
angular separation, and hence the division between resolved and unresolved sources, does not depend on the chosen passband, is commonly used by authors. Below, the legitimacy of such an approach is challenged.

\begin{figure}[h] 
\includegraphics[width=.49\textwidth]{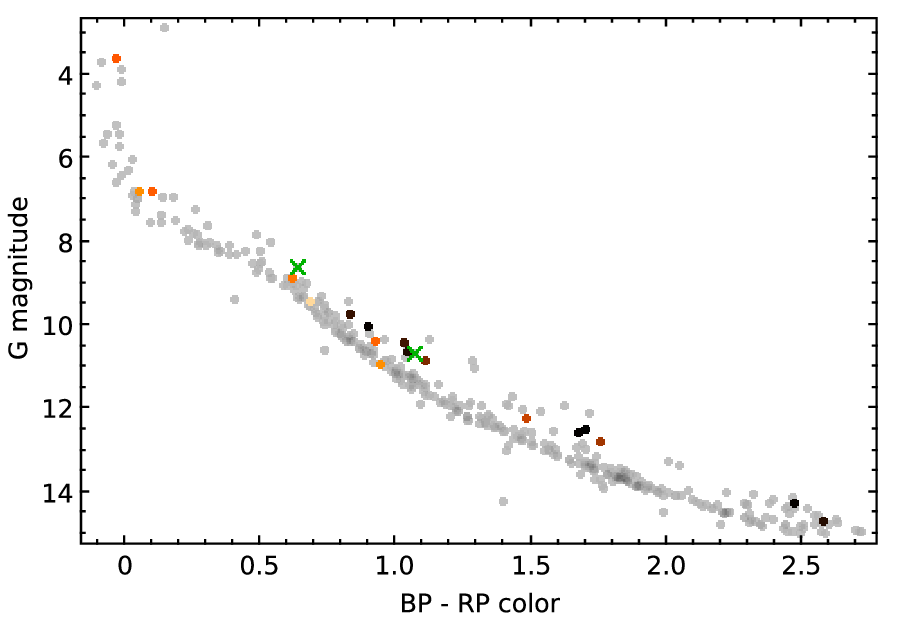}
\includegraphics[width=.49\textwidth]{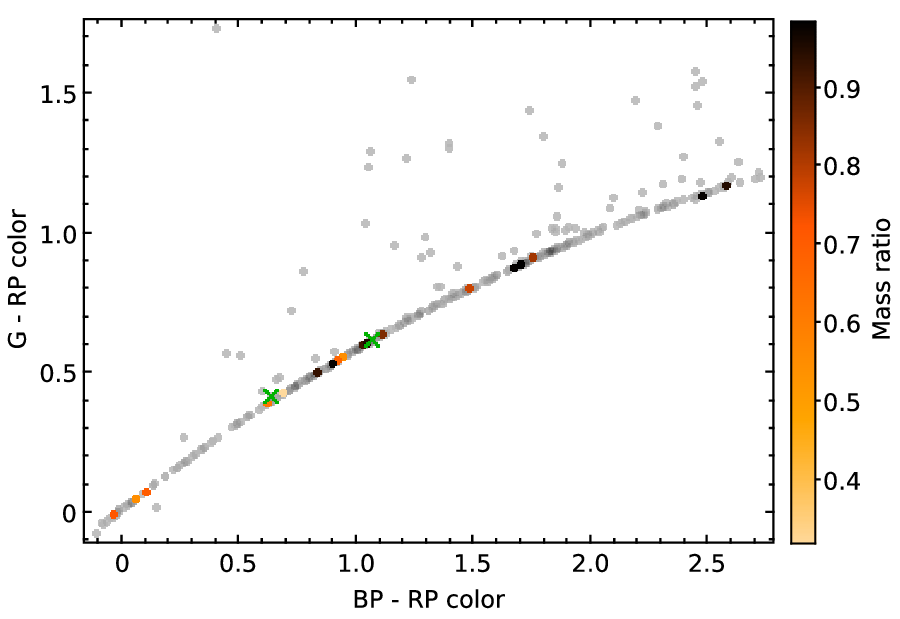}
\caption{\textbf{Left}: 18 unresolved binary systems with known mass ratios of components (color-coded) in the color-magnitude diagram among other Pleiades members. The majority of objects are double-lined spectroscopic binaries from the \cite{2021ApJ...921..117T} survey. Two faint ($G>14^{\rm mag}$) binaries were revealed by \cite{2021AJ....162..184K}. Triple-lined multiple systems are marked with a \textcolor{green}{$\times$}.  The remaining are the astrometric-spectroscopic solutions for Atlas \citep{2004A&A...425L..45Z} and V1282 Tau \citep{2020ApJ...898....2T}, which have $\rho<0.05^{\prime\prime}$. Binaries with large $q$ and triple systems are shifted above the main sequence. \textbf{Right}: the same unresolved systems in the color-color plot show no displacement.}
\label{fig:SB2} 
\end{figure}

\begin{figure}[h] 
\includegraphics[width=.49\textwidth]{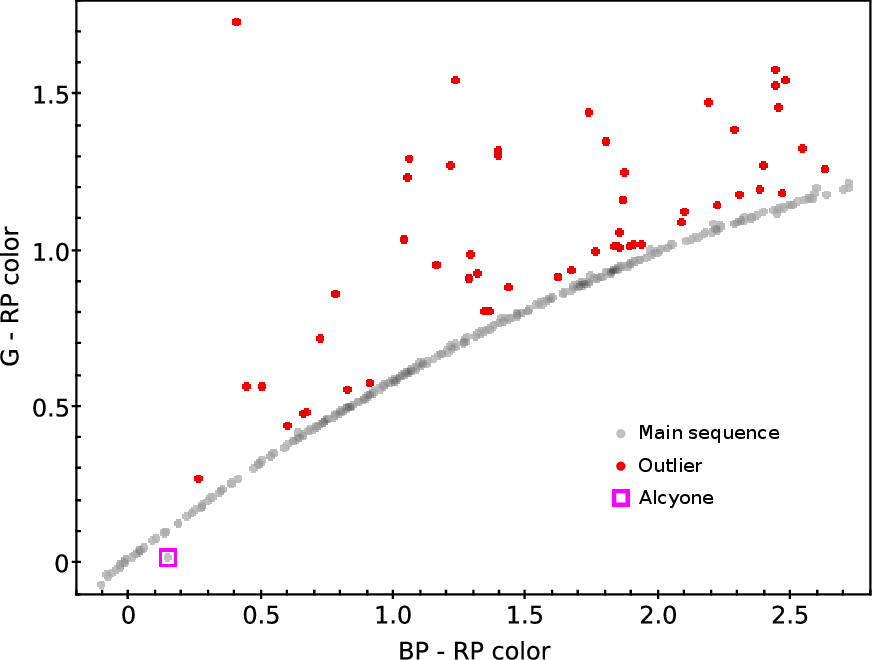}
\includegraphics[width=.49\textwidth]{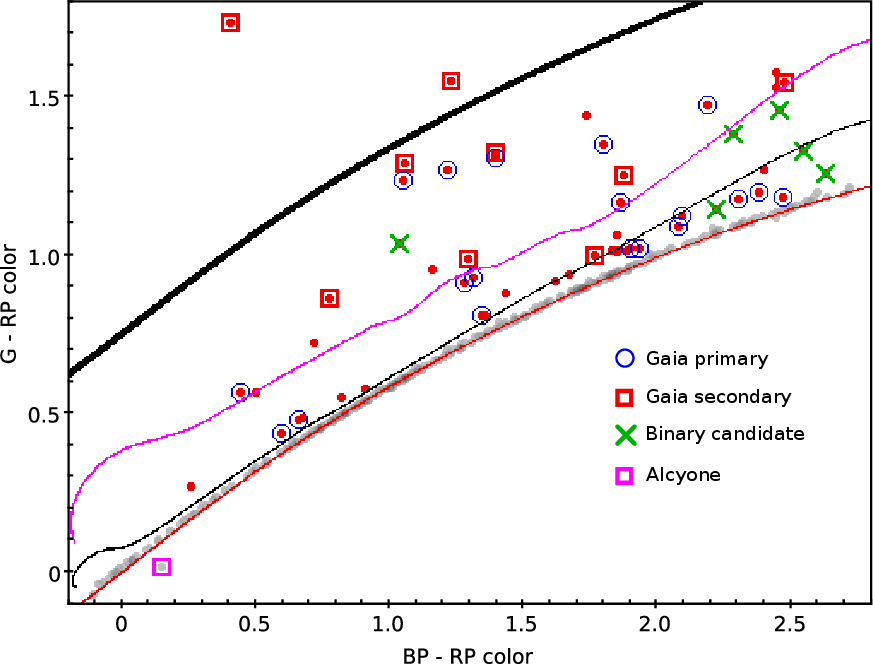}
\caption{
Color-color diagram for the Pleiades stars. The $G-R_p$ color is chosen for illustration; a similar pattern is observed with $B_p-G$ on the ordinate axis.  \textbf{Left}: 54 \textcolor{red}{outlying sources} selected for further analysis. \textbf{Right panel}: \textcolor{red}{Red line}: expected relation for single stars from the PARSEC isochrone (Section \ref{isochrones}, Figure \ref{fig:PARSEC}). The unreddened (absorption neglected) colors are chosen for a better fit with observational data. Other lines show binary populations with different mass ratios. Thin black: $q=0.5$, \textcolor{violet}{violet}: $q=0.75$, thick black: $q=1$. Importantly, $G$ flux is calculated for the primary star alone, while $B_p$ and $R_p$ are for the entire binary system. \textcolor{blue}{$\bigcirc$} and \textcolor{red}{$\square$} mark components of resolved Gaia DR3 pairs; \textcolor{green}{$\times$} indicate sources with unconfirmed multiplicity. Bare \textcolor{red}{$\cdot$} denote unresolved Gaia sources whose multiplicity is known from ground-based observations (Table \ref{tab:outliers}).}
\label{fig:G-RP} 
\end{figure}

The Pleiades cluster analysis provides an opportunity to complement Gaia data with long-term studies of the binary stars. Known double-lined spectroscopic systems are used to probe the behavior of unresolved systems in Figure \ref{fig:SB2}. As expected, spectroscopic binaries are shifted above the main sequence, which is populated by single stars, with the offset depending on the mass ratio. When the $G-R_p$ color is plotted instead of the $G$ magnitude, these systems fall on the main sequence. Simulations with the theoretical isochrone (Section \ref{isochrones}) confirm that populations of single and unresolved binary stars should be indistinguishable in the color-color plot. Yet, at least 54 entries stand out in Figure \ref{fig:G-RP}. Analysis shows that the vast majority of these sources are resolved binary stars with angular separation within 2 arcsec. Differences in field size for flux measurements in $B_p$, $R_p$, and $G$ passbands likely explain the outliers. Thus, the reported $G$ magnitude refers to an individual star, while the $B_p$ and $R_p$ values represent the total system's flux.

Depending on the angular separation and magnitude contrast, one or both components of a binary system may appear in Gaia DR3 as separate sources. Nearly half of the outlying entries in Figure \ref{fig:G-RP} belong to Gaia-resolved pairs, with 9 entries being secondary components within their systems. Among these, the faint ($G_2=13.83^{\rm mag}$) neighbor ($\rho=8.8^{\prime\prime}$) of Asterope ($G_1=5.75^{\rm mag}$) stands out, as its measured flux in $B_p$ and $R_p$ is probably affected by the very bright star in close vicinity. Another example is  HD 23247 ($G_1=8.97^{\rm mag}$, $G_2=14.21^{\rm mag}$ at $\rho=3.8^{\prime\prime}$). The other pairs have $\rho<2^{\prime\prime}$, smaller than the 3.5 $\times$ 2.1 arcsec window used for $B_p$ and $R_p$  flux measurements \citep{2021A&A...649A...3R}. At low $\rho$, only components of similar brightness can appear separately in Gaia DR3. Among cluster members selected in Section \ref{sample}, only five pairs have $\rho<1^{\prime\prime}$, all of which appear in Figure \ref{fig:G-RP} as outliers. 

For further validation, a mock population is generated in such a way that the $G$ magnitude represents the flux from the primary star alone, while $B_p$ and $R_p$ refer to the combined magnitude of the binary system. This constrains the mass ratio using the reported $B_p-R_p$, $B_p-G$, and $G-R_p$ colors. Generated populations with $q=$ 0.5, 0.75, and 1.0 are shown in the color-color diagram (Figure \ref{fig:G-RP}). For Gaia-resolved systems, $q$ inferred from the observed colors is compared with the value obtained from the known $G$ magnitudes of the components (Section \ref{mass}). The color-derived $q$ tends to be lower than that based on $G$ magnitudes, especially for wider systems with $\rho \sim 2^{\prime\prime}$, possibly because the secondary star contributes to $B_p$ and $R_p$ fluxes only partially. For binaries with $\rho<1^{\prime\prime}$, the agreement  is reasonable (Table \ref{tab:outliers}). However, it is safer to consider the color-derived $q$ as a lower limit of the actual mass ratio.

\begin{table}
     \caption{    
    List of outlying sources in the color-color diagram (Figure \ref{fig:G-RP}). First, 5 Gaia-resolved pairs with $\rho<1^{\prime\prime}$ are shown.  Next, 20 binaries whose multiplicity is confirmed by other observations are listed. Two mentioned systems are known as multiple: both companions of WDS 03520+2440 are resolved, while WDS 03500+2351C is a single-lined spectroscopic system with a 6.5 year period \citep{2021ApJ...921..117T}. The table concludes with yet unresolved binary candidates. References: Gaia \citep{2023A&A...674A...1G}, WDS \citep{2001AJ....122.3466M} (\textit{Precise Last Only} version is used), MH09  \citep{2009ApJS..181...62M}.}
    \label{tab:outliers}
    \centering
    
     \resizebox{\textwidth}{!}{\begin{tabular}{|c|c|c|c|c|c|c|c|c|c|}
        \hline
         Designation& RUWE&Primary $G$  & Contrast &Passband& Reference & Separation&contrast-  & color- \\
         \hline
         &&mag&$\Delta$ mag& &&arcsec&\multicolumn{2}{|c|}{derived $q$}
         \\ \hline
         \multicolumn{8}{|c|}{\bf Part I. Resolved Gaia DR3 pairs}\\ \hline
         WDS 03494+2456& 9.95& 11.24 & 0.06 & $G$&Gaia & 0.28 & 0.99 & $>$ 0.95\\
        \hline
         	2MASS J03431902+2226572&22.253&11.79&0.26& $G$&Gaia&0.54&0.96&$>$ 0.94 \\
          \hline
         
        WDS 03461+2452& 7.545 & 11.95 & 1.57 &  $G$&Gaia& 0.86 & 0.75 & $>$ 0.73  \\ \hline
        WDS 03466+2421& 3.342 & 12.44 & 0.30&  $G$&Gaia& 0.89 & 0.96 & $>$ 0.93 \\ \hline
         V353 Tau&1.309&13.85&3.38& $G$&Gaia&0.80&0.36 $^{+0.02}_{-0.01}$ &$>$ 0.34 \\ \hline
        \multicolumn{9}{|c|}{\bf Part II. Resolved binary stars}\\ \hline
        WDS 03456+2420& 11.507& 7.25 & 1.6 & $V$ & WDS & 0.15 & 0.68 $\pm$ 0.03 & $>$ 0.59 \\ \hline
        WDS 03471+2449 &12.858& 8.48 & 2.0 & $V$ & WDS & 0.62 & 0.69 $\pm$ 0.03& $>$ 0.74 \\ \hline
        WDS 03486+2411& two-pam& 9.01 & 1.1 & $V$ & WDS & 0.19 & 0.82 $\pm$ 0.03 & $>$ 0.57 \\ \hline
        WDS 03457+2454 &two-pam& 9.42 & \multicolumn{2}{|c|}{Lunar occultation}&WDS &0.15& & $>$ 0.55 \\ \hline
        WDS 03481+2409 &9.686& 9.44 & 0.9 & $V$ & WDS & 0.51 & 0.85 $\pm$ 0.03 & $>$ 0.78 \\ \hline
        WDS 03500+2351 C &11.596& 10.23 & 2.07 & $J$ & B97 & SB1 \& 0.53&triple& $>$ 0.53  \\ \hline

        WDS 03509+2350 & two-pam& 11.41 & 0.26 & $K$ & MH09 & 0.10 & 0.91 $\pm$ 0.03 & $>$ 0.81 \\ \hline
        WDS 03520+2440 &11.672& 11.73 & 0.34 \& 1.34 &  $J$ & B97 & 0.11 \& 0.6 & triple & $>$ 0.64\\ \hline

       WDS 03457+2345 &5.671& 11.93 & 2.41 & $J$ & B97 & 0.50 & 0.47 $\pm$ 0.03& $>$ 0.44 \\ \hline

       WDS 03434+2500 &2.198& 12.24 & 1.8 & $J$ & B97 & 0.73 & 0.60$^{+0.02}_{-0.03}$ & $>$ 0.54\\ \hline

       WDS 03445+2410 &5.186& 12.94 & 1.84 & $J$ & B97 & 0.50 & 0.53$^{+0.04}_{-0.03}$ & $>$ 0.48\\ \hline
WDS 03471+2343&3.969&13.32 &0.58&$I$&H18&0.60&0.90$\pm$0.01 & $>0.93$\\ \hline
WDS 03467+2456&5.573&13.44&2.58&$I$&H18&0.44&0.47$\pm$0.03&$>$ 0.48 \\ \hline
WDS 03419+2327&1.849&13.49&1.86&$I$&H18&0.90&0.64$\pm$0.03 & $>$ 0.60\\ \hline
WDS 03465+2407&7.651&13.54&1.8&$J$&B97&0.32&0.51 $^{+0.03}_{-0.04}$ &$>$ 0.45\\ \hline
WDS 03496+2327&12.566&13.60&2.55&$I$&H18&0.65&0.46 $^{+0.04}_{-0.03}$ &$>$ 0.39\\ \hline
WDS 03492+2333&14.578&13.68&1.77 &$J$ &B97 & 0.69&0.50 $\pm$ 0.04 &$>$ 0.45\\ \hline
WDS 03437+2434&10.146&14.65 & 1.67 & $I$ & H18   & 0.78 & 0.54$^{+0.03}_{-0.02}$ & $>$ 0.47 \\ \hline
WDS 03491+2344&1.508&14.78&1.21 & $I$ & H18&0.45&0.66 $\pm$ 0.03 &$>$ 0.80\\ \hline
WDS 03502+2400&two-pam&14.78&0.92&$I$ & H18 & 0.59 &0.74 $^{+0.03}_{-0.02}$  & $>$ 0.75 \\ \hline
        
        \multicolumn{9}{|c|}{\bf Part III. Suspected binary candidates }\\ \hline
         	2MASS J03434748+2312424& two-pam & 10.77 & \multicolumn{5}{|c|}{Unresolved by B97 \Bouvier} & $>$ 0.87 \\ \hline
         V346 Tau &19.453&14.00 & \multicolumn{5}{|c|}{Unresolved by H18 \citep{2018AJ....155...51H}} & $>$ 0.36\\ \hline
          	2MASS J03402464+2249002 &two-pam&14.27& \multicolumn{5}{|c|}{} & $>$ 0.71\\ \hline
         V336 Tau &two-pam&14.51 & \multicolumn{5}{|c|}{Unconfirmed memberdhip status} & $>$ 0.68\\ \hline
         V467 Tau &two-pam&14.59 & \multicolumn{5}{|c|}{Unconfirmed membership status, unresolved by H18} & $>$ 0.47\\ \hline
        QV Tau&two-pam&14.66 & \multicolumn{5}{|c|}{Unresolved by H18 \citep{2018AJ....155...51H}} & $>$ 0.32\\ \hline

    \end{tabular}}
   
\end{table}

Besides Gaia-resolved pairs, 20 outlying sources are known as visual binaries with subarcsecond separation. The observations for these systems are inhomogeneous and come from the data sets listed in Table \ref{tab:outliers}. The reported magnitude contrast is not very reliable for some systems, biasing the $q$ estimation. Still, contrast- and color-derived mass ratios show decent agreement. The largest separation among these binaries is 0.9 arcsec. Pairs with larger $\rho$ are successfully resolved by Gaia unless the secondary star is too faint to affect the combined color of the binary. 

The lowest reported separation among outliers is around 0.1 arcsec (Table \ref{tab:outliers}), with two objects having $\rho<0.18^{\prime\prime}$, which is a duplicity threshold in Gaia DR3 \citep{2021A&A...649A...2L}. However, the actual $\rho$ at the Gaia epoch J2016.0 may be larger due to orbital motion or measurement error. Systems with smaller separations (WDS 03491+2347, $\rho \sim 0.06^{\prime\prime}$; WDS 03477+2256, $\rho \sim 0.09^{\prime\prime}$; and WDS 03488+2416, $\rho \sim 0.10^{\prime\prime}$) definitely belong to the main sequence in Figure \ref{fig:SB2} together with spectroscopic binaries, indicating they are unresolved in the $G$ band. There is a considerable chance that some other binaries with $\rho\gtrsim 0.1^{\prime\prime}$ remain unresolved in the $G$ passband and therefore do not stand out in color-color diagram. Among the systems discovered by \cite{1997A&A...323..139B}, WDS 03490+2312 ($\rho=0.32^{\prime\prime}$, $\Delta J = 2.5^{\rm mag}$, $q \sim 0.5$) and WDS 03460+2344 ($\rho=0.51^{\prime\prime}$, $\Delta J=4.4^{\rm mag}$, $q<0.25$) have flux ratios too low to be revealed as binary candidates through color analysis, while all pairs with $\rho<1^{\prime\prime}$ from the \cite{2018AJ....155...51H} survey are detected. This suggests that the obtained sample of suspected binaries can be incomplete for $q<0.5$. 

Finally, six outlying sources from Figure \ref{fig:G-RP} are not reported as resolved binary stars despite previous observational attempts (Table \ref{tab:outliers}).  Given the challenges of ground-based high-angular-resolution observations and their dependence on atmospheric conditions, there is a good chance their multiplicity will be confirmed in the future.

\section{Conclusions}
\label{conclusions}

In the $0.1<\rho<1$ arcsec angular separation range, only 5 binaries (4 with $q>0.5$) are directly resolved by Gaia. The displacement in the color-color diagram (Figure \ref{fig:G-RP}), induced by different flux measurement pipelines, suggests subarcsecond multiplicity for 26 other sources. Of these, 20 (17 $\pm$ 3 have $q>0.5$) are confirmed by prior high-angular-resolution observations (Table \ref{tab:outliers}). Overall, within the projected separation and mass ratio range of $13.5<\rho<135$ AU and $q>0.5$, at least $24 \pm 3$ stars are identified as binary, indicating a fraction of 6 $\pm$ 1\%. Virtually all these stars show RUWE excess in Gaia DR3 or have two-parameter solutions, underscoring the complex imprint of binarity on RUWE, which is not limited to systems with orbital periods within a few years \citep{2024arXiv240414127C}. 

This study focuses on a narrow separation range, unlike other works that provide bulk binarity statistics, making direct comparison difficult. While multiplicity properties depend on stellar mass, the observed peak for solar neighborhood stars neatly fits the surveyed 13.5 -- 135 AU range \citep{2010ApJS..190....1R, 2019AJ....157..216W}. Given that this trend is relevant for the Pleiades and assuming a dispersion of lognormal separation distribution $\sigma_\rho=1.5$  across  $-2 < \log a \ [\rm  AU] < 4.5$ \citep{2023ASPC..534..275O}, the investigated range approximately accounts for one-fourth of the population, meaning a marginalized binary fraction of $22 \pm 4$\% for $q>0.5$ binaries or around 16\% for $q>0.6$. This extrapolation, though model-dependent and ill-conditioned, shows reasonable agreement with results from other authors: 
$22 \pm 4$\% for $q>0.6$ binaries by \cite{2023ApJS..268...30L}, $14 \pm 2$\% for $q>0.6$ by \cite{2021AJ....162..264J}, $18.8 \pm 0.5$\% for $q>0.4$ by \cite{2023AJ....166..110P} or $15 \pm 10$\% by \cite{2023A&A...672A..29C} for $q>0.6$. Some underestimation ($8\pm 1$\% for $q>0.6$ by \cite{2023A&A...675A..89D}) of multiplicity in open clusters could be anticipated because sources with high RUWE or two-parameter solutions in Gaia DR3 are often excluded from member lists under consideration.


\begin{acknowledgments}
Author is grateful to Mikhail Kovalev (Yunnan Observatories), Anton Seleznev (Ural Federal University), Lyudmila Mashonkina, Yury Pakhomov, Sergey Vereshchagin (INASAN), Alexey Rastorguev, Boris Safonov and Ivan Strakhov (Sternberg Astronomical Institute) for fruitful discussions on this work. \textcolor{black}{ Comments and suggestions from the referee  allowed to improve the paper.} This study was carried out using the equipment bought with the funds of the Program of the Development of M.V. Lomonosov Moscow State University. 
\end{acknowledgments}

\bibliography{sample631}{}
\bibliographystyle{aasjournal}



\end{document}